\documentclass{elsart}

\usepackage{graphicx}

\begin{document}
\begin{frontmatter}
\title {Comparison of Radiation Damage
in Lead Tungstate Crystals under Pion and Gamma Irradiation}
\author[IHEP]{V.A.~Batarin},
\author[FNAL]{J.~Butler},
\author[NAN]{T.Y.~Chen},
\author[IHEP]{A.M.~Davidenko},
\author[IHEP]{A.A.~Derevschikov},
\author[IHEP]{Y.M.~Goncharenko},
\author[IHEP]{V.N.~Grishin},
\author[IHEP]{V.A.~Kachanov},
\author[IHEP]{V.Y.~Khodyrev},
\author[IHEP]{A.S.~Konstantinov},
\author[IHEP]{V.I.~Kravtsov},
\author[UMN]{Y.~Kubota},
\author[IHEP]{V.S.~Lukanin},
\author[IHEP]{Y.A.~Matulenko},
\author[IHEP]{Y.M.~Melnick},
\author[IHEP]{A.P.~Meschanin},
\author[IHEP]{N.E.~Mikhalin},
\author[IHEP]{N.G.~Minaev},
\author[IHEP]{V.V.~Mochalov},
\author[IHEP]{D.A.~Morozov},
\author[IHEP]{L.V.~Nogach},
\author[IHEP]{A.V.~Ryazantsev},
\author[IHEP]{P.A.~Semenov},
\author[IHEP]{V.K.~Semenov},
\author[IHEP]{K.E.~Shestermanov},
\author[IHEP]{L.F.~Soloviev},
\author[SYR]{S.~Stone},
\author[IHEP]{A.V.~Uzunian\thanksref{addr}},
\thanks[addr]{corresponding author, email: uzunian@sirius.ihep.su}
\author[IHEP]{A.N.~Vasiliev},
\author[IHEP]{A.E.~Yakutin},
\author[FNAL]{J.~Yarba}
\collab{The BTeV electromagnetic calorimeter group}
\date{\today}

\address[IHEP]{Institute for High Energy Physics, Protvino, Russia}
\address[FNAL]{Fermilab, Batavia, IL 60510, U.S.A.}
\address[NAN]{Nanjing University, Nanjing,China}
\address[UMN]{University of Minnesota, Minneapolis, MN 55455, U.S.A.}
\address[SYR]{Syracuse University, Syracuse, NY 13244-1130, U.S.A.}

\begin{abstract}
  Studies of the radiation hardness of lead tungstate crystals produced
  by the Bogoroditsk Techno-Chemical Plant in Russia and
  the Shanghai Institute of Ceramics in China have been
  carried out at IHEP, Protvino. The crystals were
  irradiated by a 40 GeV pion beam. After full recovery,
  the same crystals were irradiated using a $^{137}$Cs $\gamma$-ray source.
  The dose rate profiles along the crystal length were observed to be quite similar.
  We compare the effects of the two types of radiation
  on the crystals light output.

\end{abstract}
\end{frontmatter}

\section{Introduction}
Lead tungstate ($PbWO_4$) crystals are planned for use in as the
absorptive media in the electromagnetic calorimeters in some
collider experiments. The crystals will be irradiated by high
energy particles and accumulate significant absorbed doses, up to
a several Mrad. The radiation hardness of PbWO$_4$ crystals has
been studied by the CMS group using radioactive sources and
electron beams \cite{cmsn} and by the BTeV group using both high
energy pion and electron beams. A mixture of pion and electron
beams  provides a particle environment similar to that expected
when running the BTeV experiment
 \cite{btev,nim3}.

The goal of this study is to compare the radiation damage of
crystals using two types of irradiation, either a high energy pion
beam or a $^{137}$Cs radioactive source. If the effects of these
two types of radiation are similar,
 we could use source measurements to classify the radiation
 resistance of each crystal in the BTeV like environment.
 This would be considerably easier than using particle beams
 for this measurement.
 Assessing the radiation resistance, is a crucial aspect of crystal quality assurance
 needed before assembling
the electromagnetic calorimeter.
 However, until now, there has not been a comparative study of PbWO$_4$ crystal
radiation damage under hadron beam and gamma source irradiation.

The expected dose distribution in the BTeV crystals is shown
 in Table~\ref{tab:summary} for a collider luminosity
 of $2\times 10^{32}$ cm$^{-2}$s$^{-1}$ and a running time of $10^7$ s.
 \begin{table}[htb]
    \caption{Fraction of BTeV crystals with given absorbed doses
and dose rates estimated at the maximum of the dose profiles
 inside the crystals (100 rad = 1 Gy) }
    \label{tab:summary}
    \begin{center}
      \begin{tabular}{|c|c|c|} \hline
Relative    & Absorbed dose & Dose rate  \\
number (\%)  & (krad/year)  &  (rad/h)    \\ \hline \hline

33         & 0.3 - 2     &  0.11 - 0.72    \\ \hline
27         & 2 - 5     &  0.72 - 1.8    \\ \hline
12         & 5 - 10     &  1.8 - 3.6    \\ \hline
16         & 10 - 50    &  3.6 - 18    \\ \hline
6.2         & 50 - 100     &  18 - 36    \\ \hline
3.2         & 100 - 200     &  36 - 72    \\ \hline
2         & 200 - 500     &  72 - 180    \\ \hline
0.4         & 500 - 1000     &  180 - 360    \\ \hline
0.2         & 1000 - 2000     &  360 - 720    \\ \hline
      \end{tabular}
    \end{center}
  \end{table}

 Since the BTeV calorimeter starts close to the beam-line, about
 15 cm away, and extends to a radius of 160 cm, the amount of radiation
 varies greatly. Near the beam-line a small fraction
 of the crystals, $\sim$1\% have doses of about $10^4$ Gy (1 Mrad),
 while 90\% of the crystals have doses less than 0.1 Mrad.
 The latter corresponds to a dose rate of 36 rad/h.
 About 70\% of the crystals have dose rates $<$ 4 rad/h.\\

 We used three crystals produced by the Bogoroditsk Techno-Chemical Plant
in Russia (denoted as CMS-2442, CMS-2443, Bogo-2313 ) and three more
crystals produced by
the Shanghai Institute of Ceramics in China
(denoted as Shan-T16, Shan-T19, Shan-T20 ). All the crystals except
those which are marked as CMS were produced at the end of 2000.
The CMS crystals were produced in 2001.
 The dimensions of the crystals were $\sim 27\times 27$ mm$^2$
 in cross section and $220$ mm in length.
The CMS crystals had the same length but were tapered; the
dimension at the back is $29\times 29$ mm$^2$.
 Each crystal was wrapped using 170 $\mu$m thick Tyvek.\\

The crystals  were irradiated by a 40 GeV $ \pi ^- $ beam, four of
them in April 2001 and the other two, the CMS crystals, in April
2002.
After the recovery process at room temperature the crystals were
irradiated by $^{137}$Cs $\gamma$-source in May and June 2003.
 We provided very similar
irradiation conditions,  in terms of dose rates and absorbed
doses, for both cases. Two types of measurements were done on each
crystal both before and after the gamma irradiation: a) light
output (LO) measurements using a radioactive source for crystal
excitation, and
 b) crystal transparency measurements using a spectrophotometer.

\section{Irradiation by pions}

Pion irradiation was carried out at the IHEP (Protvino) 70 GeV
accelerator. The size of the 40 GeV $\pi^-$  beam was 8 cm
horizontally and 6 cm vertically, $i. e.$ 90\% of the beam was
contained within these dimensions. The beam was present in 1.2
second interval of the full accelerator
 cycle of 9 sec.
Six crystals
were irradiated with a dose rate ranging from 30 to 60 rad/h.
The crystals light output signals were monitored using the minimum
ionizing particles (MIP) in pion beam running and from separate
low-intensity calibration runs in electron beam running. LeCroy
2285 15-bit integrating ADC's were used to measure the signal
using a 150 ns integration time.
 The Protvino testbeam setup is described in detail elsewhere
 \cite{nim3,nim1,nim2}.

\section{Irradiation using a $^{137}$Cs source}

\begin{figure}
\centering
\includegraphics[width=95mm]{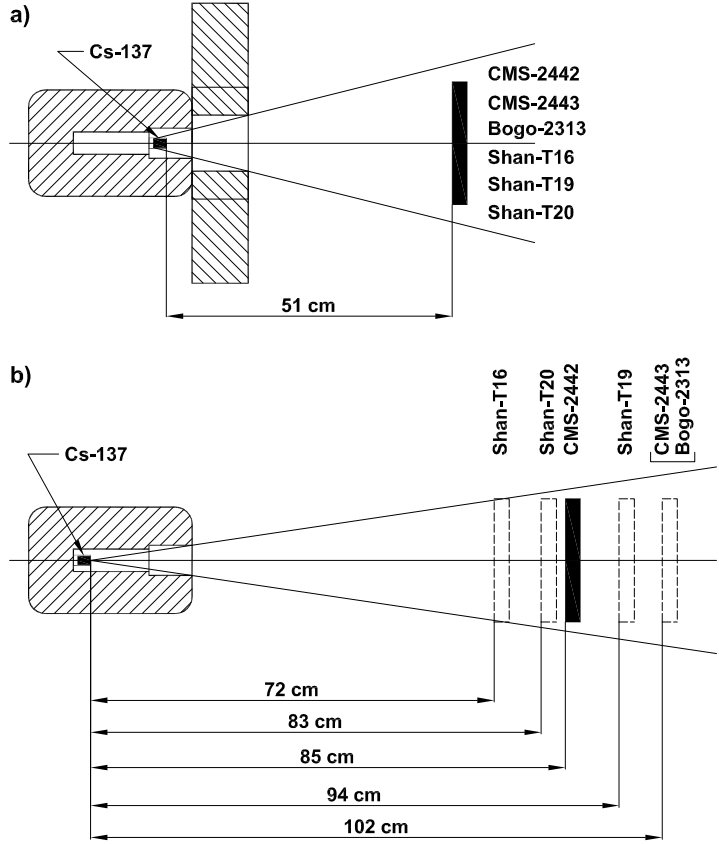}
\caption{Setup of the crystal gamma irradiation at
the dose rate of 110 rad/h (a), and the dose rates of 30-60 rad/h (b).}
\label{fig:gsetup}
\end{figure}

We used a radioactive source $^{137}$Cs, that emits 661 KeV
photons and had an activity
 of  $5\cdot 10^{12}$ Bq to irradiate the crystals.
The geometrical setup for each crystal position is shown in
Fig.~\ref{fig:gsetup}. Each crystal was irradiated separately. A
commercial dosimeter, DKS-AT1123, was used to measure
 the dose rate (in air) in proximity to the crystals.
The accuracy is better than 25\% \cite{http}.

\begin{figure}
\centering
\includegraphics[width=90mm]{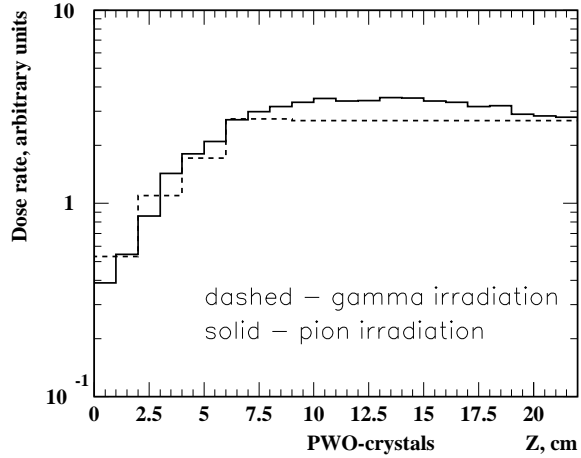}
\caption{Longitudinal dose rate profiles for crystal
 under pion and gamma irradiation(Z near 22 cm corresponds to the
phototube front).}
\label{fig:dprof}
\end{figure}

\begin{figure}
\centering
\includegraphics[width=110mm]{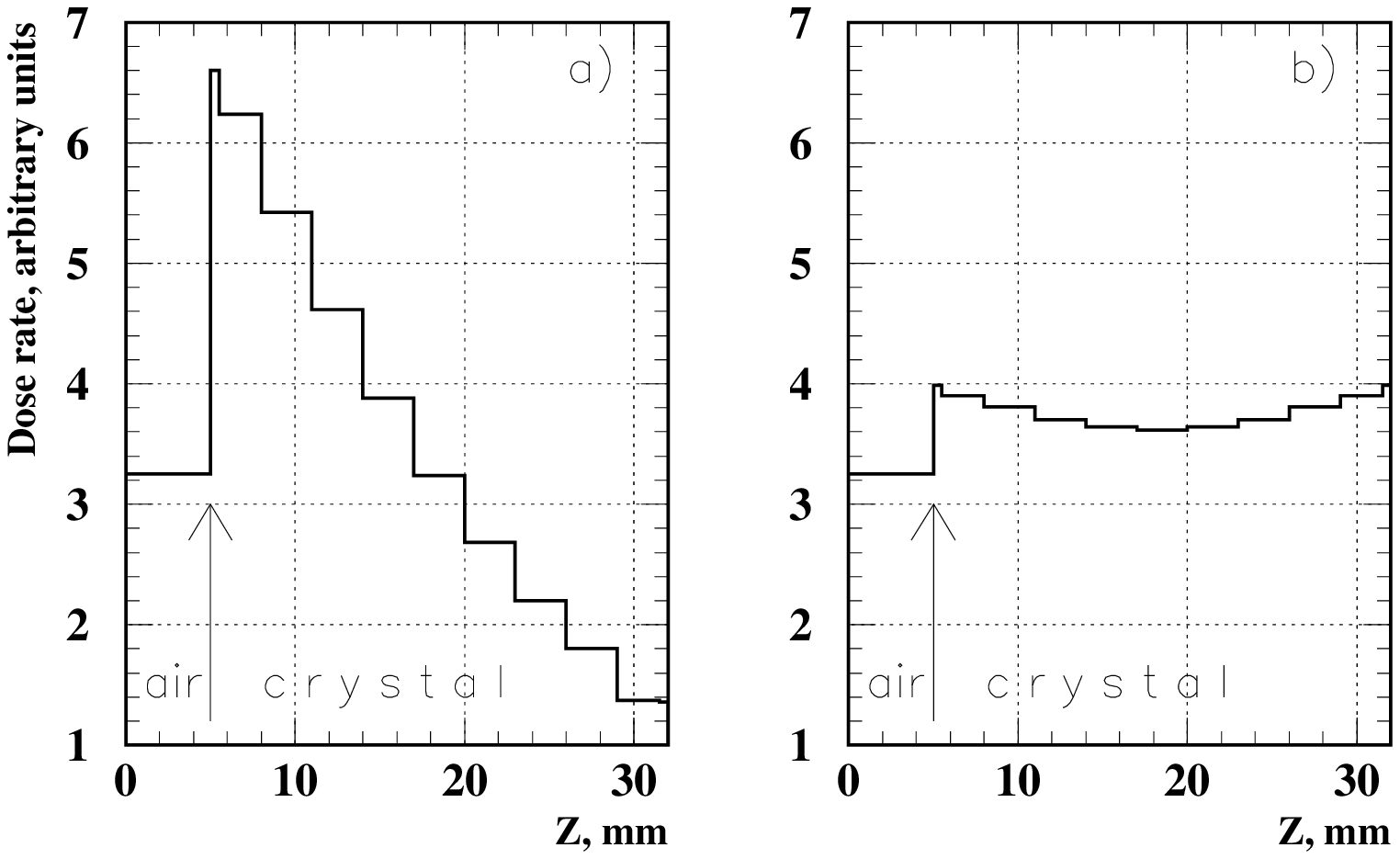}
\caption{ Transverse dose rate profiles for crystal under
$^{137}Cs$ gamma irradiation:
 a) side irradiation without crystal rotation, b) side
irradiation with
 crystal rotation. Dose estimations were made in both ``air" and
``crystal," as indicated.}
\label{fig:cs137prof}
\end{figure}
To modify the spatial dose distribution from the source in order
 to make it as similar as possible to pion beam, we used a
 lead screen placed between the source and the crystals and also rotated
 the crystals along their longitudinal axes at a rate of 2 turns per minute.
The lead screen was made using four lead plates with varying
 dimensions \cite{leadscreen}.
The dose rate distributions along the length of crystal are compared
 for both the pion and source irradiation in Fig.~\ref{fig:dprof}
 using the lead screen and rotating the crystals.
 We see that the profiles are quite similar.
The rotation is even more important for insuring a uniform
 irradiation transverse to the crystal axis when using the source.
 In Fig.~\ref{fig:cs137prof} we show the transverse radiation
 profiles for case (a) where the crystal was not rotated and for case
 (b) where it was rotated.
 Since pion irradiation uniformly illuminates the crystals
 in transverse direction,
 we use the crystal rotation for all the source results in this paper.
The MARS and ROZ-6 codes were used for simulations of the dose
profiles in crystal under pion and gamma irradiation respectively
\cite{nim3,roz}.

 Two sets of gamma irradiation were carried out. One of them was done
 using dose rates of (30-60) rad/h, and the second one at the dose rate of
110 rad/h for each crystal.
 The crystals were allowed to recover during two weeks,  at room temperature,
 between the first and the second sets of irradiation.
 After each exposition the crystal was removed from the place of irradiation
 to dedicated test stands for light output and transmittance measurements.
 This procedure usually took about 1 to 1.5 hours.

\section{Light output and transmittance measurements}

A stand with a radioactive source was used to study changes in the
crystal light output due to radiation damage. Scintillation light
was detected by a XP2020Q photomultiplier tube (PMT). The PMT
output was connected to a 20 kOhm load. The mean current through
this load was measured by a 11-bit ADC read out by a computer. The
PMT high voltage value was of 1.5 kV. The PMT nonlinearity in the
direct current mode was less than 1\%
 in the range of 1.2 kV to 2 kV.
The dark current was  $\sim 2\cdot 10^{-3}$ relative to the
average signal level. The background current, mostly due to direct
PMT irradiation by the source was $\sim 5\cdot 10^{-3}$ of the
average signal in the worst instance. A $^{137}$Cs source ($7\cdot
10^8$ Bq) was installed inside a cylindrical lead collimator, 30
mm in diameter and 40 mm in height with aperture diameter of 5.3
mm. The crystals were wrapped by a Tyvek, except for one end which
was attached to the PMT without optical grease. The source was
moved longitudinally parallel to the central line of the crystal
with steps of 2 cm. The distance between the source and the
crystal surface was 5 mm.
 The measurements of the average direct current
 were provided in the same way for each of the 9 position along the crystal.
The mean value over these 9 measurements gives us a value of a
 light output signal of each crystal.
The light output measurements were monitored using a non-irradiated
 reference crystal.
 Performing measurements on non-irradiated crystals, we
estimate the systematic error to be at most 3\% of the relative
 light output degradation.
The temperature was measured during irradiation and during light
output monitoring.
 The temperature variations  were $\sim$ 1 degree on average, and
up to 2 degrees in the worst case. There were no corrections for
the temperature variations. The full systematic errors of the
monitoring measurements include the effects of the temperature
differences.

To measure the light transmittance of each crystal, a commercial
spectrophotometer SF-26 was used in the wave length region from
340 up to 700 nm with a steps of 10 nm. The spectrophotometer
light spot was 5 mm in diameter with a negligible angular
dispersion.
 The measurement accuracy at each point was 1\%.

\section{Results and discussion}

\begin{figure}
\centering
\includegraphics[width=100mm,height=100mm]{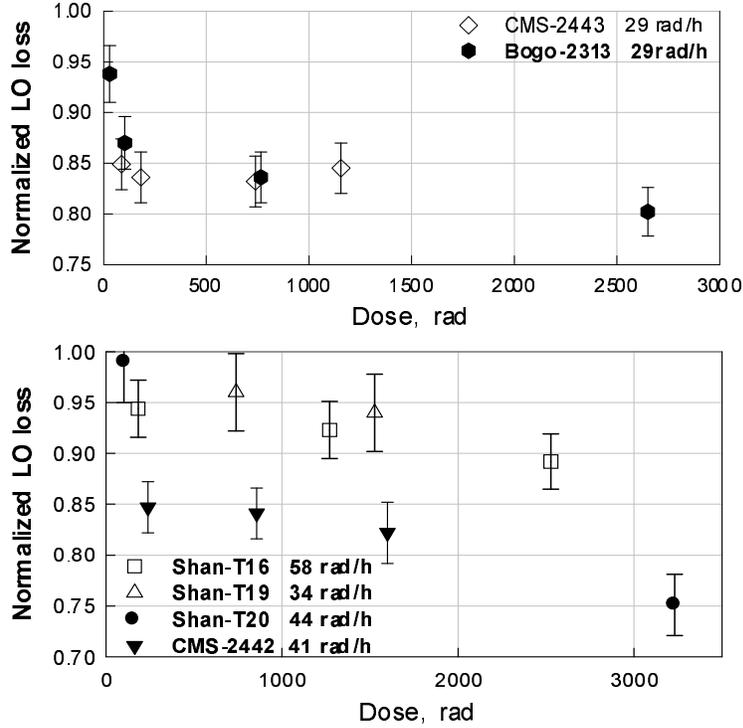}
\caption{Light output loss for six crystals versus absorbed dose
 under gamma irradiation at a fixed dose rate for each crystal.
 There were
several measurements for the two crystals, CMS-2443 and Bogo-2313
(upper part).
The results of measurements for other four crystals (bottom part)
are consistent with the first two, but were taken at longer intervals.}
\label{fig:Cs137_lin}
\end{figure}

\begin{figure}
\centering
\includegraphics[width=80mm,height=60mm]{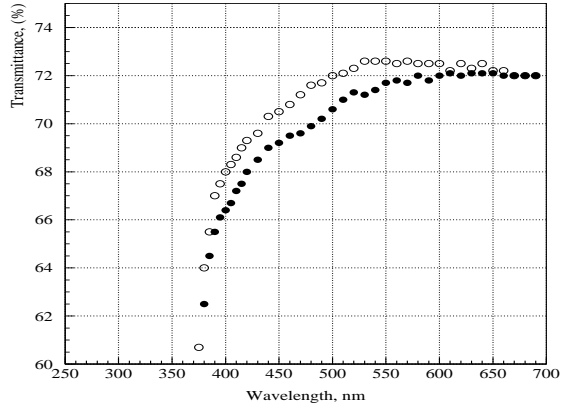}
\caption{ Transmittance of the CMS-2442 crystal before (open circles)
 and after (filled circles) gamma irradiation at 41 rad/h dose rate
 and 4.4 krad of absorbed dose.}
\label{fig:phspec}
\end{figure}
The dependence of a signal loss on the accumulated dose (at a
fixed dose rate) for six crystals irradiated by $^{137}$Cs source
is shown
 in Fig.~\ref{fig:Cs137_lin}.
 Two crystals lost
 10\% of their signal or less under the source irradiation and
 four other crystals lost (15-25)\%.
The light loss in transmittance curves at wavelengths of (420-550)
nm measured with the use of the spectrophotometer is one order of
magnitude less. For example crystal CMS-2442 showed losses between
2-2.5\%
 (see Fig.~\ref{fig:phspec}).
This difference might be due to different optical paths
in crystal taken by the injected spectrophotometer light
as compared with the scintillation light.
From now on we will show only the results from
 the radioactive source measurements which are more sensitive
 than the spectrophotometer measurements
 in showing the effects of radiation.

It has been shown that PbWO$_4$ transmission damage occurs in the
crystal when valence electrons are trapped in metastable states
around crystal defects.
 Thus, the irradiation of PbWO$_4$ crystals creates
so called color centers which reduce the light attenuation length.
When the rate of color centers production (proportional to the
dose rate) equals the natural recovery rate, the crystal light
output will reach
 a saturation level \cite{ann}.
 Fig.~\ref{fig:Cs137_lin} shows that the signal degrades relatively
rapidly with the absorbed dose up to (100-200) rad and then degrades
at a significantly slower rate until the saturation level is reached.
The light loss has a tendency to exhibit saturation when the dose rate
is kept at a constant level.
Each crystal demonstrates a different level of light loss
 in the saturated state.
We found that the saturation level is reliably reached after
 $\sim$10 hours of continuous irradiation at the fixed dose rate.
We did not observe a significant difference in radiation hardness of
the crystals from different manufacturers.
\begin{figure}
\centering
\includegraphics[width=110mm]{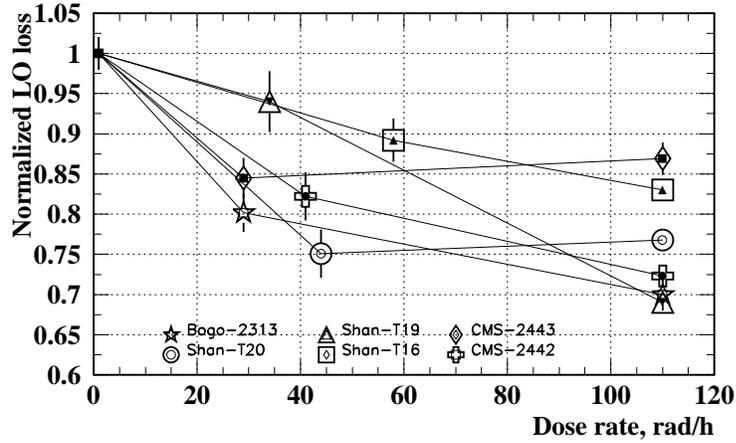}
\caption{Dependence of the normalized light output loss
 (in state of saturation)
 on a dose rates for six crystals for gamma irradiation using
 $^{137}$Cs source.}
\label{fig:gdrate}
\end{figure}

\begin{figure}
\centering
\includegraphics[width=110mm]{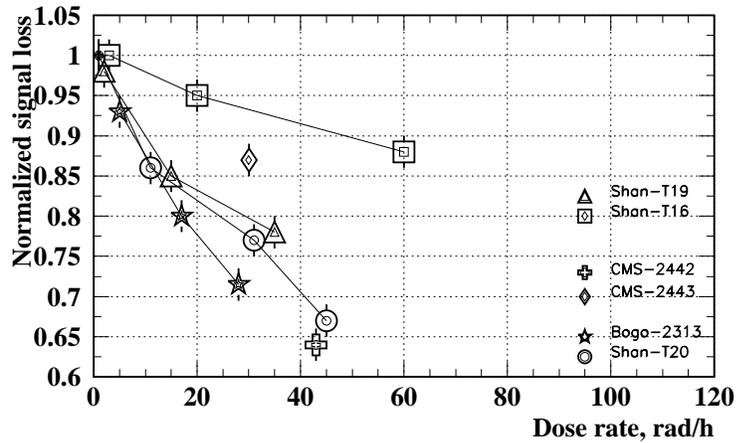}
\caption{ Dependence of the normalized signal loss (in state of saturation)
 on a dose rates for 40 GeV pion irradiation. }
\label{fig:pidrate}
\end{figure}

The second round of gamma irradiation was done at a dose rate
 of 110 rad/h for each crystal, after two weeks of recovery
 at room temperature.
Because we irradiated the crystals with 110 rad/h gamma-source
for a long time, $\sim$(60-70) hours,
we believe that we reached the signal loss saturation level for
each crystal at this high dose rate.
 The dependence of the normalized signal loss (in state of saturation)
 on the dose rate for the six crystals under gamma irradiation
 is shown in Fig.~\ref{fig:gdrate}.

The normalized signal loss versus of the absorbed dose under 40
GeV pion irradiation for the same crystals is presented in
Fig.~\ref{fig:pidrate}. The points on the plot represent the level
of signal loss
  after saturation was reached at the specified fixed dose rates.
 The rate of the
signal degradation is practically the same for both irradiation
procedures up to the dose rates of 60 rad/h.
Using gamma irradiation at a dose rate
of 110 rad/h does not lower the saturation level of some of the crystals,
 while others are affected (see Fig.~\ref{fig:gdrate}).
This is an indication that the relative difference of the crystal
signal losses demonstrated at dose rates of 30-60 rad/h might
change at larger dose rates.

\begin{figure}
\centering
\includegraphics[width=100mm,height=80mm]{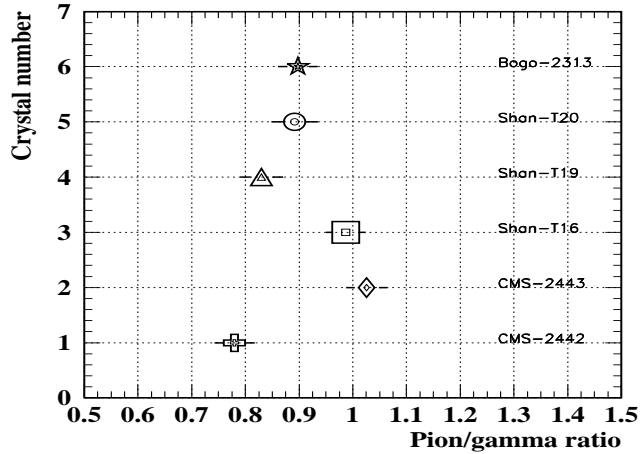}
\caption{ Ratios of two signal losses at their
saturation levels, under pion and under gamma irradiation
for the same dose rates (30-60 rad/h) for six crystals.
}
\label{fig:pigamrat}
\end{figure}

Ratios of pion/gamma signal losses at saturation, caused by
irradiation at the same dose rates, for the six crystals are shown
in Fig.~\ref{fig:pigamrat}.
 We see that these
ratios are close to each other and are in the region
between 0.8 and 1.
Thus we can conclude that pion and gamma irradiation at the same
dose rates affect lead tungstate crystals in a similar fashion.

We intend to continue our studies; we want to carry out crystal
radiation studies for significantly larger numbers of crystals
while making continuous light output measurements during the gamma
irradiation process.

\section{Summary}
  We have observed that the relative signal loss using
40 GeV pion irradiation and using $^{137}$Cs radioactive source
irradiation looks similar at the dose rate of (30-60) rad/h.

 We have shown that the gamma-source can be used to evaluate the radiation tolerance of
 PbWO$_4$ crystals. It takes approximately 10 hours of irradiation
 at fixed dose rates  to reach a saturation
 plateau for each crystal.
The dose rate can be chosen in the range of (60-100) rad/h because the
most of the crystals will be suffering from the lower
dose rates at the BTeV experiment.

\section{Acknowledgments}
This work was partially supported by the U.S. National Science
Foundation and the Department of Energy as well as The Russian
Foundation for Basic Research grant 02-02-39008.
    We would like to thank the  management of IHEP Radiation Research
Department for providing us a $^{137}$Cs radioactive source for
our gamma irradiation studies, as well as the PHENIX group at IHEP
(Protvino) for providing us both hardware and software needed
 for light-output measurements.
We thank O.~Sumaneev for calculation of the dose profiles in the
crystals using gamma irradiation.


\begin{thebibliography}{99}

\bibitem {cmsn} CMS, The Electromagnetic Calorimeter Project Technical
                  Design Report, CERN/LHCC 97-33, CMS TDR 4 (1997); \\
 S.Baccaro et al.,{ Further understanding of PbWO$_4$
scintillator characteristics and their optimization},
CERN, CMS NOTE 2000/002 (2000);\\
S. Gascon-Shotkin, {Recent Developments in Crystal
 Calorimeters (featuring the CMS PbWO$_4$ Electromagnetic Calorimeter)
},
CERN, CMS CR 2003/001 (2003);\\
G.Davies et al., { A study of the monitoring of radiation damage
 to CMS ECAL crystals, performed at X5-GIF},CERN, CMS NOTE 2000/020 (2000);\\
H.F.Chen et al.,{ Measurements of PbWO$_4$ crystals behavior under
 irradiation}, CERN CMS CR 1999/027 (1999).

 \bibitem {btev} A.~Uzunian, A.~Vasiliev, J.~Yarba,
{ First results on simulation of radiation environment
    at BTeV electromagnetic calorimeter},
IHEP preprint 2001-24, Protvino 2001.

\bibitem{nim3} V.A.Batarin et al., Study of Radiation Damage in Lead
Tungstate Crystals Using Intense High Energy Beams,
e-Print ArXiv hep-ex/0210011;
Nucl. Instr. and Meth. A 512(2003) 488.

\bibitem{nim1}    V.A.~Batarin {\sl et al.},
                  Development of a Momentum Determined Electron Beam in
the 1-45 GeV Range, e-Print Archive
hep-ex/0208012; Nucl. Instr. and Meth. A 510(2003) 211.

\bibitem{nim2}    V.A.~Batarin {\sl et al.},
                  Precision Measurement of Energy and Position Resolutions
of the BTeV Electromagnetic Calorimeter Prototype, e-Print Archive
hep-ex/0209055 ; Nucl. Instr. and Meth. A 510(2003) 248.

\bibitem{http}  http://www.atomtex.com/.

\bibitem{leadscreen}
The lead screen was formed of plates with dimensions
 (thickness, length, height)xmm$^3$: 19x20x40, 12x20x50,
 7.5x20x50 and 2x30x50.

\bibitem{roz} A.V~Averin, A.M.~Voloschenko, E.P.~Kondratenko{\sl et al.},
The ROZ-6 One-Dimensional Discrete Ordinates Neutrons, Gamma-Rays
and Charged Particles Transport Code,\\
Proc. Int. Topical Meeting on Advaces in Mathematics, Computations
and Reactor Physics, Pittsburgh USA, 1991, vol.5.

\bibitem{ann} A.A~Annenkov, M.V.~Korzhik, P.~Lecoq,
{Lead tungstate scintillation material},
Nucl. Instr. and Meth. A 490(2002) 30.




\end{thebibliography}
\end{document}